\documentstyle[12pt]{article}
\begin{document}
\begin{titlepage}
\hspace*{\fill}{IMSc-96/06/19}
\vspace*{\fill}
\begin{center}
{\Large \bf The asymptotic behaviour of $F_L$ in the double scaling
limit}\\[1cm]
Tapobrata Sarkar and Rahul
Basu\footnote{email:sarkar,rahul@imsc.ernet.in}\\
{\em The Institute of Mathematical Sciences, Madras 600 113, India}
\end{center}
\vspace{2cm}
\begin{abstract}
In the kinematic region of small $x$ and large $Q^2$ in deep inelastic
scattering, presently being explored by HERA, 
we present an analysis of the evolution of the longitudinal structure
function $F_L^{p}(x, Q^2)$ in the double scaling limit of Ball and Forte.
We fit the evolution to a $1/x^\lambda$ behaviour and extract the
value of $\lambda$. We also study the behaviour of $R=F_2/2xF_1 -1$.
We present comparisons of both $F_L$ and $R$ with the corresponding 
MRS fits in this region of $x$ and $Q^2$. 
\end{abstract}
\vspace*{\fill}
\end{titlepage}

The recent spurt of activity in the physics of low $x$ QCD has been
guided by the electron-proton scattering measurements at HERA which
have studied the proton structure function $F_2^{p}(x,Q^2)$ in the very 
low $x$ region, $x\leq 10^{-3}$. The steep rise of $F_2$ at these small 
values of $x$ have given rise to a flurry of theoretical activity in an
attempt to understand the nature of this rise through 'standard' QCD.

There have essentially been two approaches to the study of the HERA
data. In order to study the effects of summing $\alpha_s\ln {1\over
x}$ unaccompanied by $\ln Q^2$, the preferred approach has been to
study the BFKL equation which generates a singular $x^{-\lambda}$
behaviour for the unintegrated gluon distribution $f(x,k_T^2)$, with
$\lambda={\bar \alpha_s} 4\ln 2$ (for fixed, not running $\alpha_s$).
For running $\alpha_s$, $\lambda\simeq 0.5.$ \cite{bfkl}. From this an
asymptotic form for $F_i(i=2,L)$ can be inferred.

The other distinct approach is to attempt to describe the data through
a 'standard' Altarelli-Parisi (or DGLAP) evolution equation to the
next-to-leading order approximation. Here again, the data imply a
steep gluon distribution with the gluon density rising sharply as $x$
decreases, even for comparatively low values of $Q^2$. Ball and Forte
\cite{bf}, in a series of papers, have used this approach to exhibit
the scaling properties at low $x$, generated by QCD effects and have
shown that the HERA measurement of $F_2(x,Q^2)$ is well explained by
this approach.

We describe their result very briefly. They show that at sufficiently
small $x$ and large $t=\ln\frac{Q^2}{\Lambda^2}$, the structure
function $F_2$ exhibits double scaling in the variables
\begin{equation}
\sigma\equiv\sqrt{\ln\frac{x}{x_0}\ln\frac{t}{t_0}}\ \ ;
\rho\equiv \sqrt{\frac{\ln x_0/x}{\ln t/t_0}}
\end{equation}
where the starting scale for $Q_{0}^{2}$ in  
$t_0\equiv \ln \frac{Q_0^2}{\Lambda}$ can be
just a little more than $Q_0^2=1 GeV^2$, and $\Lambda= \Lambda_{QCD}$. 
Double asymptotic scaling
results from the use of the operater product expansion and the
renormalisation group at leading (and next-to-leading) order, and
predicts the rise of $F_2$ on the basis of purely perturbative QCD
evolution.

The asymptotic behaviour in this approach (to leading order) has the
form
\begin{eqnarray}
F_2^p(\sigma,\rho)&\sim& Nf(\frac{\gamma}{\rho})\frac{\gamma}{\rho}
\frac{1}{\sqrt{\gamma\rho}}exp[2\gamma\rho-\delta(\frac{\sigma}{\rho})]
\nonumber \\
&&\times [1+O({1\over\sigma})]
\end{eqnarray}
where 
$$
\gamma\equiv 2\sqrt{\frac{N_c}{\beta_0}} \ \; \beta_0=11-{2\over
3}n_f
$$
and
$$
\delta\equiv (1+\frac{2n_f}{11N_c^2})(1-\frac{2n_f}{11N_c})
$$
with $f$ an unknown distribution that depends on the starting value
of the distribution.

In this paper, we concentrate on the Ball and Forte \cite{bf}
approach to QCD evolution to study the other independent proton structure
function -- the longitudinal structure function for the 
proton, $F_L^{p}(x,Q^2)$. $F_L$
has not yet been measured at HERA and it will of course be very
interesting to see how the data compares with theory.

Before we embark on the calculation, we would like to make a few
comments about these different approaches. As is well known, in the
traditional DGLAP or Ball-Forte analysis, there are two unknowns {\em
viz.} $F_2(x,Q^2)$ and $g(x,Q^2)$ (the gluon distribution) and these
are determined by two measurements {\em viz.} a direct measurement of
$F_2$ (from the cross section) and the $Q^2$ variation of $F_2$ which
gives $g(x,Q^2)$. These two measurements fix these two unknown
quantities exactly since there are {\em no} free parameters in this
approach that need to be fitted. In contrast, in the BFKL approach, due
to the resummation of the $\ln {1\over x}$ leading to evolution also
in $x$ (by some exponent), there is a third parameter which needs to
be fitted. Since there have been only two measurements till now,
this parameter, in the BFKL approach, can only be fixed with a third
measuremnent, for example, that of $F_L$. This is in striking
contrast to the Ball-Forte DGLAP approach where there are no
parameters that can be tuned in order to fit any future measurement
of $F_L(x,Q^2)$. A theoretical analysis of the behaviour of $F_L$ and
subsequent comparison with experiment is, therefore, the need of the
hour. 
We now present our analysis of $F_L(x,Q^2)$.
\par
In perturbative QCD, the nucleon structure functions are
determined in terms of the quark and gluon distribution functions which
in turn follow the DGLAP equation,
\begin{equation}
{d{\bf G} \over dt}~=~{\alpha_{s} \over 2 \pi}{\bf P} \otimes {\bf G}
\left(Q^{2} \right)
\end{equation}
where,
\begin{center}
${\bf P}~=~
\left( \begin{array}{cc}
P_{qq}   &2n_{f}P_{qg} \\
P_{gq}        &P_{gg} \\
\end{array} \right) $
~~~~and ~~~~$ G \left(x,Q^{2} \right)~=~ 
\left( \begin{array}{c}
q_{s} \\
g     \\
\end{array} \right)
%\pmatrix{q_{s}\cr g}
$
\end{center}
%\end{equation}
are the splitting function and the parton distribution matrices
respectively.
The nonsinglet quark distribution follows a similar equation but we
shall ignore it as the non singlet contribution to the longitudinal 
structure function is negligible.
\par
The above equation $(3)$ can be solved by using standard Mellin
transformation techniques \cite{bf} and in the asymptotic limit $\sigma 
\rightarrow \infty $ 
yields the result $(2)$ for the structure
function $F_{2}$.

Note however, that in the standard case, the quark
distribution functions are defined in terms of the structure functions
$F_{2}$, {\it viz.}, 
\begin{equation}
F_{2}(x,Q^{2})~=~\left({1 \over n_{f}} \sum_{i=1}^{n}e_{i}^{2} \right)
x\left(q_{s}(x,t)~+~q_{NS}(x,t) \right).
\end{equation}
We could, however, start by using $F_{1}$ as the defining structure 
function instead of $F_{2}$ and then the definition of the former would
yield 
\begin{equation}
F_{1}(x,Q^{2})~=~{1 \over 2}\sum_{i=1}^{n_{f}}e_{q_{i}^{2}}
\left( q_{i}^{(1)}(x,Q^{2})~+~{\bar q}_{i}^{(1)}(x,Q^{2}) \right)
\end{equation}
which can be expressed in terms of the singlet and non singlet
distributions as in the usual case, and where the superscript $(1)$
denotes the fact that the distributions have been defined with respect
to the structure function $F_{1}$.
If we take the quark and gluon distributions defined with respect to 
$F_{2}$ as the standard distribution, we can compute $q^{(1)}$ and 
${\bar q^{(1)}}$ with respect to these and the result is \cite{field}
\begin{equation}
q^{(1)}(x,Q^{2})~=~\int_{x}^{1}{dy \over y}\left\{ q(y,Q^{2}) 
\left[ \delta(1~-~z)~+~\alpha_{s} \Delta f_{1}^{q}(z) \right]
~+~g(y,Q^{2})\alpha_{s} \Delta f_{1}^{g}(z) \right\}
\end{equation}
and a similar equation for $\bar q^{(1)}$, 
where we have 
$$z~=~{x \over y}~~~;~~~ \Delta f_{1}^{q}(z)~=~-{4z \over 
3 \pi}~~~;~~~ \Delta f_{1}^{g}~=~-{z(1~-~z) \over \pi}$$
Now if we calculate the structure function $F_{1}$, in        
terms of these quark and gluon densities, using $(5)$, the result, 
in terms of the standard parton distributions obtained from $F_{2}$, is 
given by  \cite{altarelli}

\begin{equation}
2F_{1}~=~\int_{x}^{1}{dy \over y} \left\{\sum_{i=q,\bar{q}}
q_{i}(y)e_{i}^{2}\left[ \delta \left( 1~-~z \right)~+~{\alpha_{s} \over 
2 \pi} \sigma^{T}_{q \gamma *} \left(z \right) \right]
~+~g(y) \left( \sum_{i=q} e_{i}^{2} \right) {\alpha_{s} \over 2 \pi}
\sigma^{T}_{g \gamma *} \left(z \right) \right\} 
\end{equation}
where $q_{i}$ and $g$ are as usual the quark and gluon parton densities,
$e_{i}$ is the charge of $q_{i}$ and in analogy with $F_{2}$, we have 
written the expression in terms of 
$ \sigma^{T}_{q \gamma *}$ and
$\sigma^{T}_{g \gamma *}$ which are the transverse virtual photon-quark
(antiquark) 
and the virtual photon - gluon cross sections respectively.
The longitudinal structure function  
$F_{L}~=~F_{2}~-~2xF_{1}$ can now be easily evaluated with the standard    
expression for $F_{2}$, in terms of the quark and gluon densities,
and, rewritten in terms of $F_{2}$, yields the
Altarelli-Martinelli formula, \cite{altarelli}  
\begin{equation}
F_{L}(x,Q^{2})~=~{\alpha_{s}(Q^{2}) \over 2 \pi}x^{2} \int_{x}^{1}
{dy \over y^{3}} \left\{ {8 \over 3}F_{2}(y,Q^{2})~+~
{40 \over 9}yg(y,Q^{2}) \left(1~-~{x \over y} \right) \right\}
\end{equation}
In deriving the above formula, one uses the lowest
order result for $F_{1}$, by introducing the running coupling constant,
$\alpha_{s}(Q^{2})$, and using the momentum dependent parton densities. 
This formula can be used to compute the value of $F_{L}$ for low $x$ and
large $Q^{2}$, which combined with calculated values of 
$F_{2}$ will give an estimate of $R$.
\par
In the present work, we work in the "double scaling" regime, i.e we
consider the asymptotic solutions of $F_{2}$ and $g$ in the limit 
$\sigma \rightarrow \infty$ and fixed but large values of $\rho$ where 
$\sigma$ and $\rho$ are  defined in $(1)$.

The evaluation of $F_{2}$ and the gluon distribution function $g$ and the
fit of $F_{2}$ to experimental values have been carried out in 
\cite{bf} and we have used the same best fit normailizations as in \cite{bf}. 
In the absence of any existing data for $F_{L}$,
in this kenematical regime, we have  
fixed an overall normalization for $F_{L}$ as $0.14$ in order to 
closely match the various parametrizations for the same.
Indeed, it would be possible to carry out a detailed analysis of the 
possible boundary conditions along the lines of \cite{bf}, once
experimental values for $F_{L}$ are obtained.We must mention here that 
we are interested in only the asymptotic form of the evolution of the
various functions and 
the DGLAP evolution equations cannot fix the numbers absolutely and these 
must be uniquely determined by the experimental data.
In particular, in a recent paper, De Roeck et al. \cite{deroeck}
have fitted the expression
$(2)$ for $F_{2}$ to the latest measurement of the proton structure function 
by the H$1$ collaboration at HERA. We have assumed the result of their fit,
which gives $\Lambda = \Lambda_{QCD} = 248MeV$ and $Q_{0}^{2}=1.12~GeV^{2}$. 

We have numerically evaluated equations $(7)$ and $(8)$ and subsequently 
$F_{L}$ and $R$, in the double scaling limit. 
Since HERA has not measured either $F_L$ or $R$, we have given a
comparison to the MRSA-fit \cite{mrs}. 
The data for $F_{L}$ is shown in figure $(1a)$ and $(1b)$, while that 
for $R$ is shown in fig $(2)$. The fits are shown on the same graph.
As is clear, apart from overall normalization, 
about the arbitrariness of which we have already 
commented before, the behaviour of $F_L$
and $R$ closely follows the fit.

In order to get a somewhat more quantitative handle on the data, 
we have done a fit of the  graph for $F_L$ to the form 
$a(Q^{2})x^{- \lambda (Q^{2})}$.
The result of the fit is shown in the Table. It is strikingly
clear from the table 
that $a(Q^2)$ reaches an asymptotic value of $0.0034$ and, more
interestingly, $\lambda(Q^2)$ reaches a asymptotic value of $0.55$ 
We will have more to say on this later.

A few words about our results are now in order. In this paper, we have
considered the form of the longitudinal structure function at very low
values of $x$ ($10^{-5}<x<10^{-2}$) for a wide range of $Q^2$. We have
also calculated $R$ and compared it with the present MRSA-fit.

We see clearly that the steep rise of the longitudinal structure function 
as predicted by hard QCD processes gives a behaviour of the form
$x^{-0.5}$.
This behaviour is similar to that predicted by the BFKL \cite{bfkl}
analysis. In the BFKL analysis, summing the $\alpha_s log(1/x)$ terms
unaccompanied by $log Q^2$ generates a singular $x^{-\lambda}$
behaviour for the unintegrated gluon distribution. For a running
coupling, the BFKL equation has to be solved numerically under some
reasonable assumptions, and it yields $\lambda\simeq 0.5$. (For a
somewhat detailed discussion, see, for example, \cite{kms}). This
behaviour subsequently gives a similar behaviour for $F_2$ and $F_L$.

The rise of $F_L$ is very close to that predicted by the
MRSA-fit even though the overall normalization is presumably different
and will be a function of the input distribution.

At this level of analysis, it is impossible to distinguish, purely
from measurements of $F_L$, the behaviour as predicted by BFKL and
that of the Ball-Forte analysis. Thus the role of the hard pomeron in
deciding the nature of the rise of the structure function is not at
all clear, at this level of analysis. It is possible that very precise
meaurements of $F_L$ coupled with the non-leading contribution to the  
Ball-Forte result will allow us to make a clear statement on the role
of the hard pomeron. However, if indeed, $F_L$ is seen to go as
$x^{-0.5}$ asymptotically, while experimentally $F_2$ goes as
$x^{-0.3}$, then, even though the absolute value of $F_2$ at these
values of $x$ and $Q^2$ is much larger that $F_L$, it is possible to
find some $x$ value at which $F_L > F_2$, if indeed this behaviour
persists. If that happens then since $2xF_1 = F_2 - F_L$, this would
imply a negative LHS. Since the LHS is the momentum density and hence
positive, some new physics must take over at this stage to reduce the
rise of $F_L$ to keep the RHS positive.
%Finally we have also checked the variation of $F_L$ as a function of
%$Q^2$. We find that $F_L \sim Q^4$. This is, of course expected since
%$\sigma_L \sim F_L/Q^2$ which implies $\sigma_L \rightarrow 0$ as
%$Q^2\rightarrow 0$ as it should since the longitudinal cross section
%should vanish in the real photoproduction limit. 

A measurement of the longitudinal structure function and $R$ will
shed light on the importance of hard processes in the kinematic
region being explored at HERA and will consequently give us more
information about the gluon.

\newpage
\begin{center}
{\bf \Large Table} \\[1cm]
\end{center}
\begin{center}
\begin{tabular}{|c|c|c|}
\hline
& & \\
$Q^2$ & $a(Q^2)$ & $\lambda(Q^2)$\\
& & \\ 
\hline
&& \\
20 & 0.0183 & -0.316 \\
&& \\
\hline
&& \\
50 & 0.00777 & -0.443 \\
&& \\
\hline
&& \\
100 & 0.00652 & -0.467 \\
&& \\
\hline
&& \\
400 & 0.00483 & -0.505 \\
&& \\
\hline
&& \\
800 & 0.00424 & -0.521 \\
&& \\
\hline
&& \\
1200 & 0.00395 & -0.529 \\
&& \\
\hline
&& \\
1600 & 0.00376 & -0.535 \\
&& \\
\hline
&& \\
2000 & 0.00362 & -0.539 \\
&& \\
\hline
&& \\
2400 & 0.00351 & -0.542 \\
&& \\
\hline
&& \\
2800 & 0.00342 & -0.545 \\
&& \\
\hline
\end{tabular}
\end{center}
\newpage
\noindent{\large \bf Figure Captions} \\[2cm]
1a and 1b: Plot of $F_L$ vs. $x$ for different $Q^2$. The dashed
lines are the MRS-fit shown for comparison.\\
2: Plot of $R$ vs $x$ for different $Q^2$. The dashed lines are the
MRS-fits shown for comparison.\\[2cm]
\noindent {\large \bf Table Caption} \\[2cm]
Values of $a(Q^2)$ and $\lambda(Q^2)$ as a function of $Q^2$ (in the
fit for $F_L \sim a(Q^2)x^{\lambda}$).
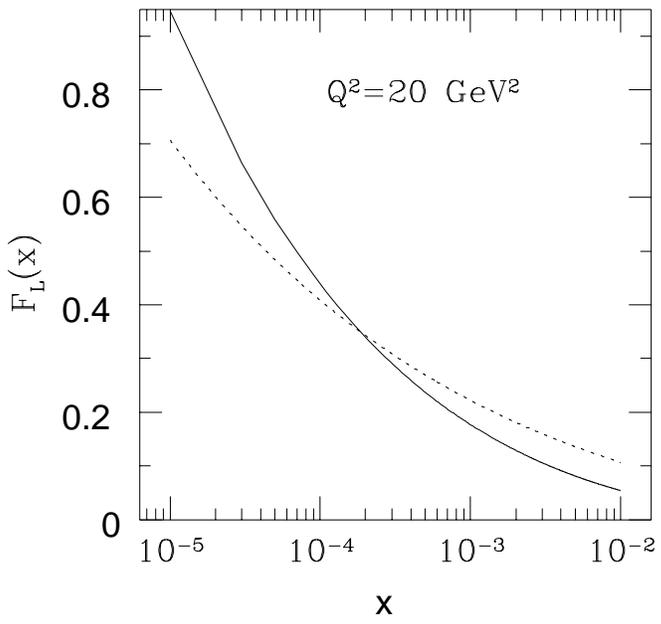
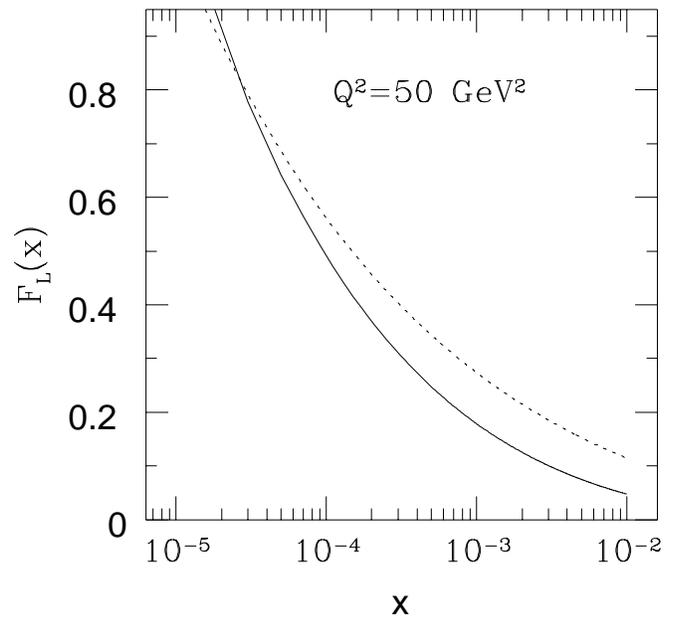
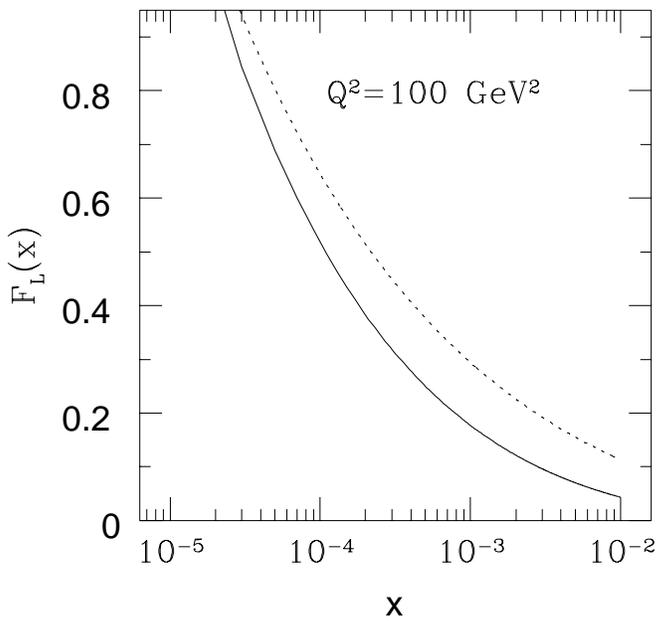
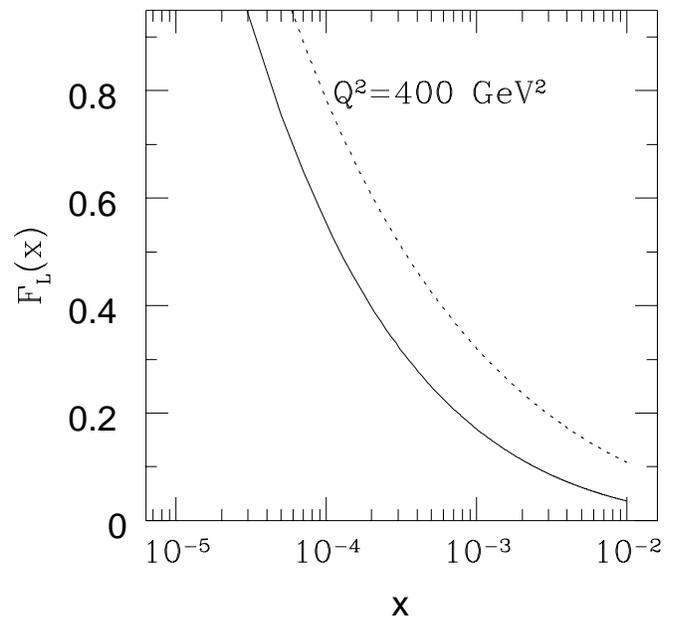
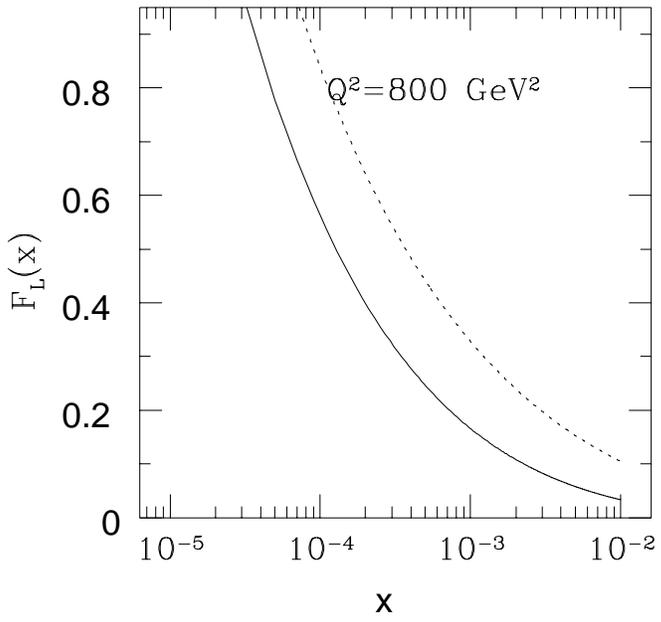
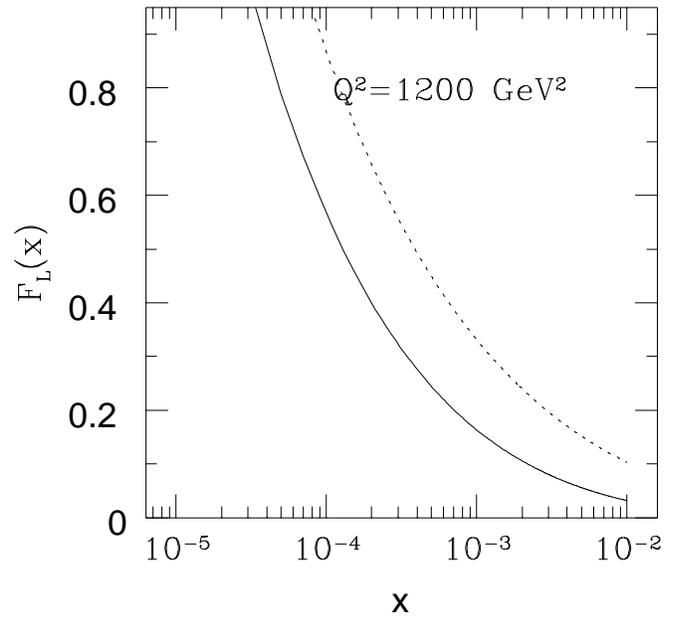
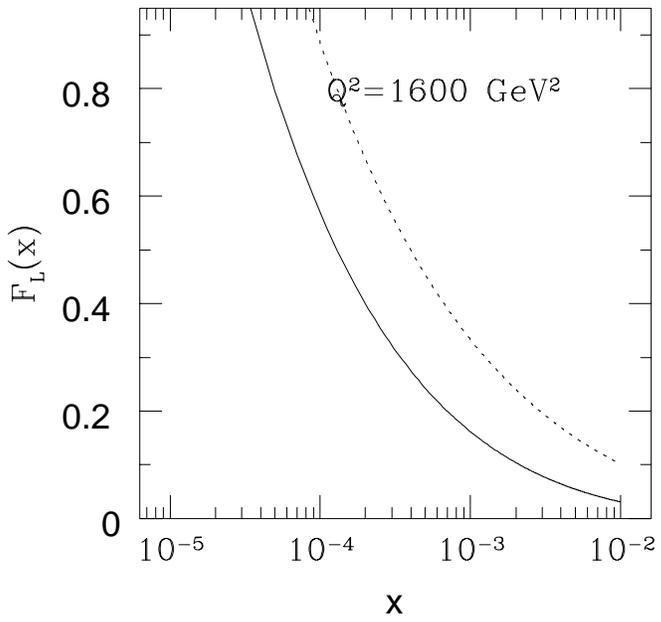
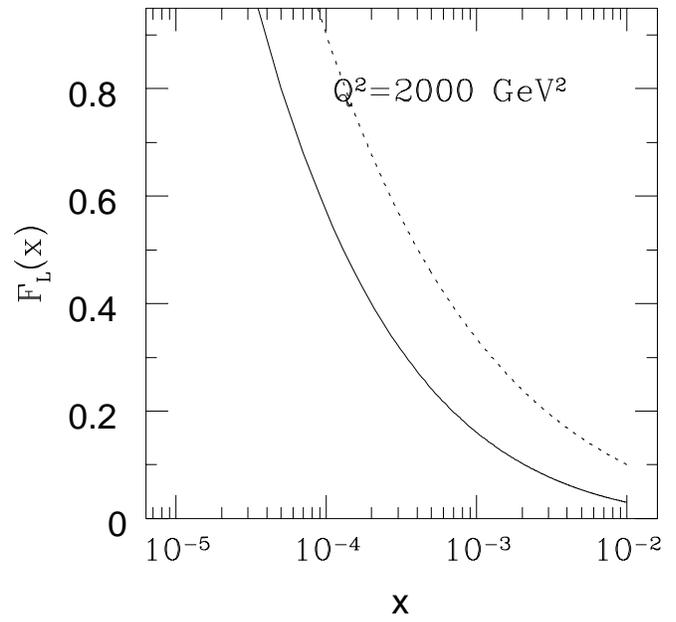
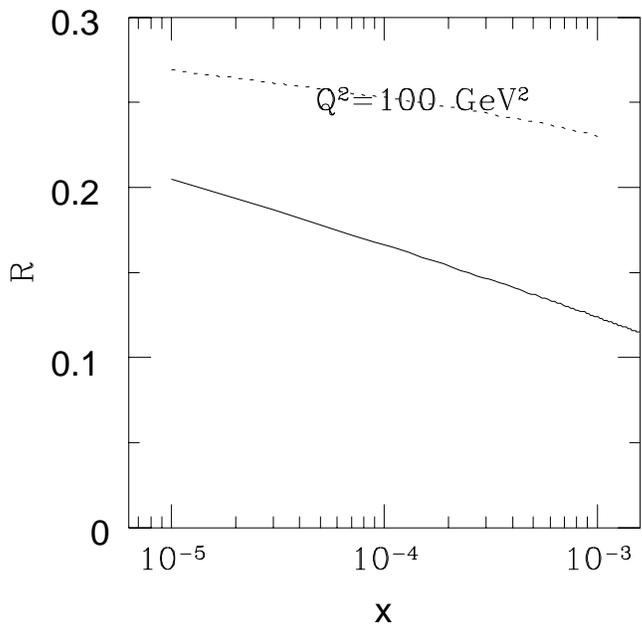
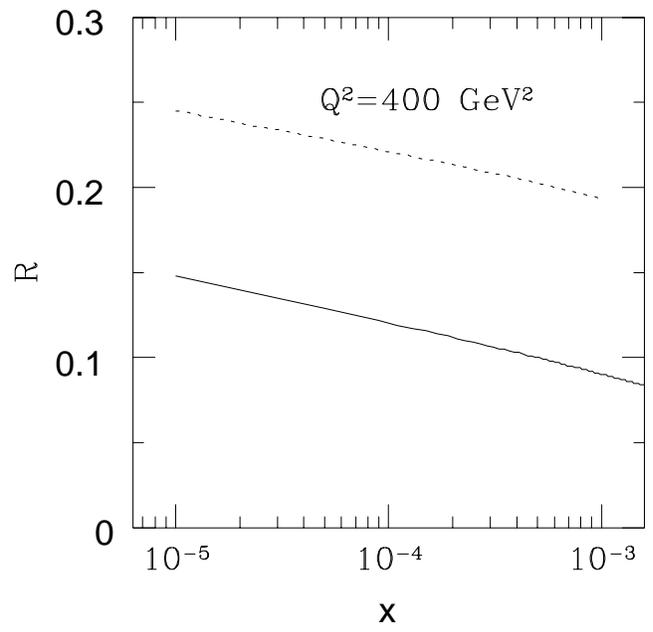
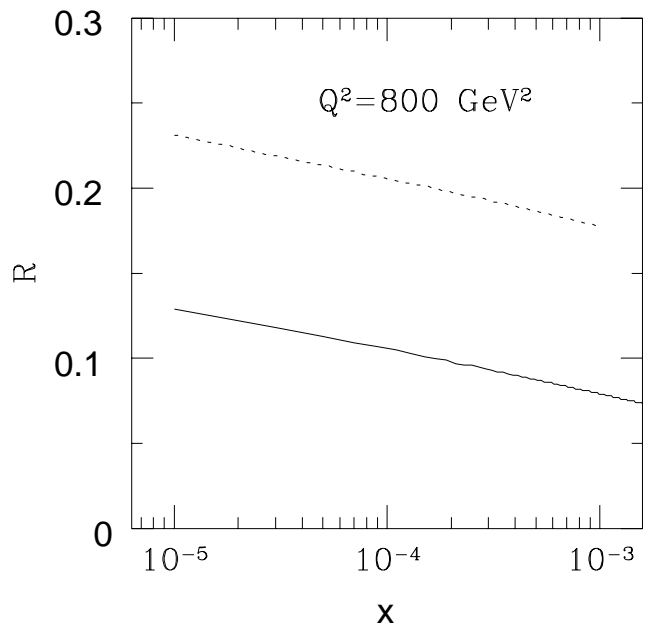
\end{document}